\documentclass[prl,reprint]{revtex4-1}
\usepackage{graphicx,verbatim,bbold,amsmath,mathtools}

\DeclarePairedDelimiter{\ceil}{\lceil}{\rceil}
\newcommand{\Q}{\mathcal Q}

\begin{document}

\title{Recovering full coherence in a qubit by measuring half of its environment}

\author{Filippo M. Miatto$^{1}$, Kevin Pich\'e$^{1}$, Thomas Brougham$^{2}$ and Robert W.~Boyd$^{1,2,3}$}
\affiliation{$^1$Dept.~of Physics, University of Ottawa, Ottawa, Canada}
\affiliation{$^2$School of Physics and Astronomy, University of Glasgow, Glasgow (UK)}
\affiliation{$^3$Institute of Optics, University of Rochester, Rochester, USA}
\date{\today}

\begin{abstract}
When quantum systems interact with the environment they lose their quantum properties, such as coherence. Quantum erasure makes it possible to restore coherence in a system by measuring its environment, but accessing the whole of it may be prohibitive: realistically one might have to concentrate only on an accessible subspace and neglect the rest. If that is the case, how good is quantum erasure?  
In this work we compute the largest coherence $\langle \mathcal C\rangle$ that we can expect to recover in a qubit, as a function of the dimension of the accessible and of the inaccessible subspaces of its environment. We then imagine the following game: we are given a uniformly random pure state of $n+1$ qubits and we are asked to compute the largest coherence that we can retrieve on one of them by optimally measuring a certain number $0\leq a\leq n$ of the others. 
We find a surprising effect around the value $a\approx n/2$: the recoverable coherence sharply transitions between 0 and 1, indicating that in order to restore full coherence on a qubit we need access to only half of its physical environment (or in terms of degrees of freedom to just the square root of them). Moreover, we find that the recoverable coherence becomes a typical property of the whole ensemble as $n$ grows. 
\end{abstract}
\maketitle

\section{Introduction}
Decoherence is a physical process that interests the scientific community from a fundamental point of view (how does the quantum-to-classical transition occur?) and also from a technical one (how can we maintain a system coherent enough throughout a quantum protocol?) \cite{Scully1989,brune1996observing, myatt2000decoherence, zurek2003decoherence,darrigo2014recovering,orieux2015experimental}. One of the techniques for restoring coherence is known as ``quantum erasure'', which consists in measuring the environment in the most appropriate basis in order to erase the information that it stores and thereby recover coherence \cite{GreenbergerYasin}. An example would be in a Young double-slit experiment where two orthogonal polarizers have been put in front of the slits and the interference fringes have disappeared. Quantum erasure (in this case with postselection) would consist of orienting a polarizer diagonally before the screen to erase the which-slit information stored in the Hilbert space of polarization (which was acting as the environment) and restore the fringes.

Quantum erasure relies on an optimal measurement of the environment of a qubit $\mathcal Q$, consisting of a number of measurement operators $\{\hat\pi_j\}$, to restore its coherence. The $j$-th measurement operator $\hat\pi_j$ projects the state of $\mathcal Q$ onto the conditional state $\hat\rho_j$, which displays a coherence $\mathcal C_j$ (defined below) \cite{englert2000quantitative}. We stress that quantum erasure does not rely on postselection \cite{menzel2012wave,bolduc2014fair,leach2014duality}, as the coherence that we maximize is the average over all the outcomes: $\langle\mathcal C\rangle=\sum_jp_j\mathcal C_k$, where $p_j$ is the probability of observing the $j$-th outcome. Recall that $\hat\rho_j$ does depend on the outcome $j$, and that the incoherent sum $\sum_jp_j\hat\rho_k$ is the same as tracing over the environment: this prevents us from measuring  coherence directly and violate the no signalling principle \cite{Miatto2015subfidelity}. It is clear that the more information one is able to erase from the environment, the more of the original coherence one is able to restore.
In our model we consider one qubit $\mathcal Q$, immersed in an environment with Hilbert space $\mathcal A\otimes \mathcal K$, where the $A$-dimensional subspace $\mathcal A$ is accessible and the $K$-dimensional subspace $\mathcal K$ is inaccessible. Therefore, we consider only measurements that span $\mathcal A$, but ignore $\mathcal K$, i.e. $\hat\pi_j=\hat\pi_j^\mathcal{A}\otimes\mathcal{\mathbb{\hat{1}}}^\mathcal{K}$. Given measurements of such type, we study how well they perform and we find that as long as $A\gtrsim K$,  one can restore almost perfect coherence on $\mathcal Q$. This phenomenon might be closely related to quantum darwinism \cite{zurek2003decoherence}.
\begin{figure}[t!]
\begin{center}
\includegraphics[width=\columnwidth]{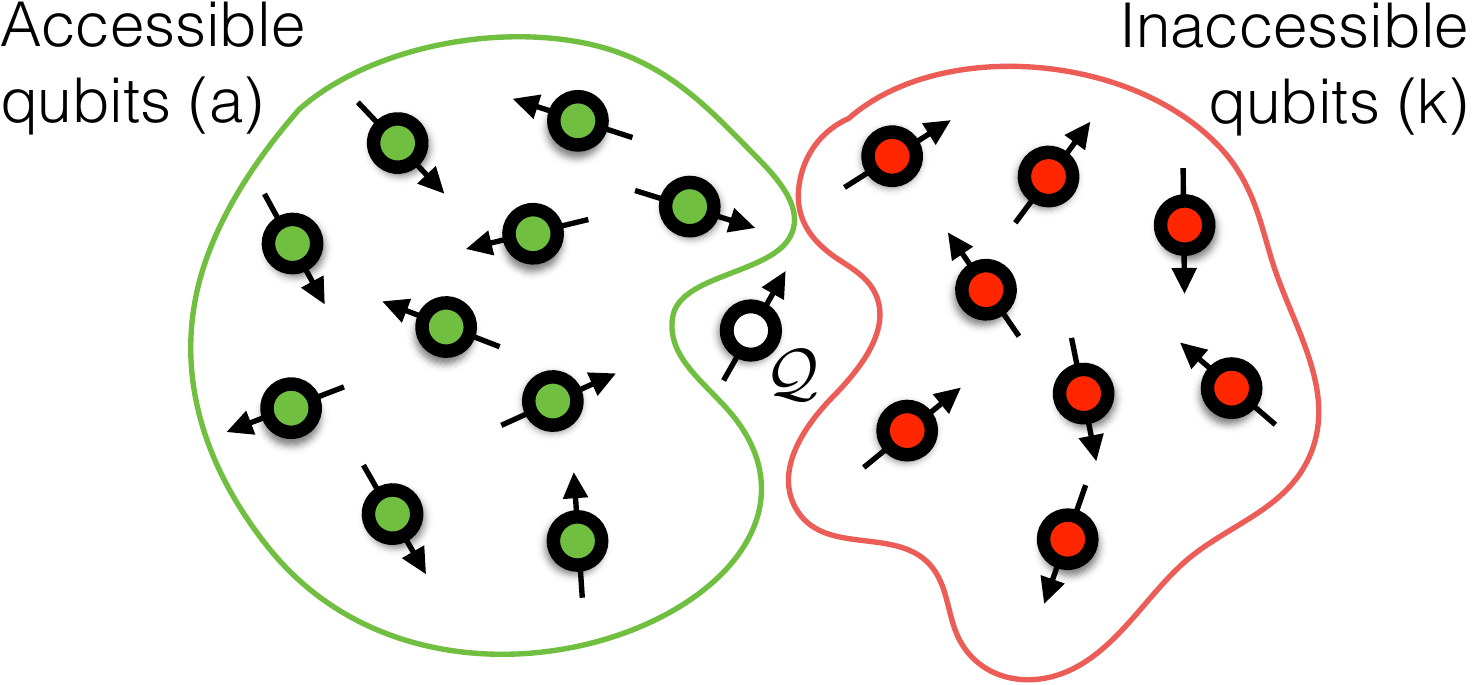}
    \caption{\label{setup}We imagine a qubit $\mathcal Q$ within an ensemble of $n$ environment qubits, where $a$ of them are accessible. The rest $k=n-a$ are inaccessible. We find that if $a\geq k$ (i.e. if we can access at least half of them) there exists an optimal measurement on the accessible qubits whose outcomes project $\mathcal Q$ onto states with coherence $\langle\mathcal C\rangle\sim1-\frac{2^k}{2^{a+2}}$, which quickly approaches 1 as $a$ increases past $n/2$.}
\end{center}
\end{figure}

The rest of this Letter is organized as follows: in the next section we define the coherence of a qubit and we show how it is influenced by a measurement on its environment. Then we split the environment into an accessible and an inaccessible part and we prove the main result. In the final section we supply examples and we show that the average coherence becomes typical as the environment grows in size.

\section{quantifying and restoring coherence}

Our first task is to identify the quantity that we are going to study, namely the coherence of $\Q$. In contrast to purity, coherence depends on the basis that we choose. In the Bloch sphere  (where the poles are the preferred basis elements $|0\rangle$ and $|1\rangle$, which we will also call ``alternatives'', implying a measurement of $\hat\sigma_z$) the coherence is the distance of the Bloch vector from the imaginary line connecting the North pole to the South pole, i.e. given a Bloch vector with coordinates $\mathbf{v}=(x,y,z)$, the coherence is $\sqrt{x^2+y^2}$, also known as ``visibility'' in view of the analogy of a qubit with a Mach-Zender interferometer (see below) \cite{Englert1996}. It is clear that if we picked a different pair of opposite points on the surface as the new North and South poles, the coherence would generally change.

We can understand the meaning of coherence by making an analogy with a photon travelling through a Mach-Zender interferometer: the two arms of the interferometer constitute the two possible alternatives and so we can represent the path of the photon with a qubit. As one varies the phase of one arm with respect to the other, the Bloch vector rotates around the $z$ axis in a circular motion of radius $\sqrt{x^2+y^2}$. This radius is precisely the visibility of the fringes at the output of the interferometer: if we had all the which-arm information, the Bloch vector would have coordinates either $(0,0,1)$ or $(0,0,-1)$, and in neither case we would see any fringes, while if the two alternatives were in a balanced coherent superposition, the Bloch vector would be on the equator and the visibility would be at its largest. Note that in order to lose the fringes we don't need to actually ``know'' which arm the photon is in, we just need such information to be ``knowable'', as in such case the qubit would be maximally entangled with some other system and its state would be maximally mixed (i.e. $\mathbf{v}=(0,0,0)$).
In terms of the density matrix of the qubit, the coherence is given by twice the absolute value of either of the off-diagonal elements.

Now, consider a qubit $\mathcal Q$ that is entangled with another system (which we call $S$ and which is not necessarily another qubit):
\begin{align}
\label{example}
|\psi\rangle=\alpha|0,s_0\rangle+\beta|1,s_1\rangle,
\end{align}
where $|s_{0}\rangle$ and $|s_{1}\rangle$ are the states of $S$ corresponding to the states of $\mathcal Q$. The density matrix of $\mathcal Q$ alone is obtained by tracing over $S$:
\begin{align}
\hat \rho_\mathcal{Q}=
\begin{pmatrix}
|\alpha|^2&\alpha^*\beta\langle s_1|s_0\rangle\\
\alpha\beta^*\langle s_0|s_1\rangle&|\beta|^2
\end{pmatrix}
\end{align} 
The coherence is therefore given by $\mathcal C=2|\alpha\beta^*\langle s_0|s_1\rangle|$, which is proportional to the overlap between the states of $S$. This happens because the more the states $|s_0\rangle$ and $|s_1\rangle$ are orthogonal, the better one can distinguish them and learn about the qubit, i.e. the more information about the alternatives of $\mathcal Q$ is stored in  $S$, which is acting as an ``environment''. This is the essence of the duality principle.

A way of erasing such information would be to measure $S$ in a basis that is unbiased with respect to $|s_0\rangle$ and $|s_1\rangle$, by way of an optimal measurement with elements $\hat\pi_j$ \cite{englert2000quantitative} defined by maximizing the mean coherence over all the probability operator measures on $S$:
\begin{align}
\label{sup}
\mathcal C&=\sup_{\{\hat\pi_j\}\in\mathrm{POM}(S)}\sum_j|2\alpha\beta^*\langle s_0|\hat\pi_j|s_1\rangle|\\
&=2\mathrm{Tr}\bigl|\alpha\beta^*|s_1\rangle\langle s_0|\bigr|
\label{tracenorm}
\end{align}
where $\mathrm{Tr}|x|$ is the trace norm of $x$.
In this way, regardless of the outcome, we would not learn which of the states $|s_0\rangle$ or $|s_1\rangle$ the environment is in, and after such measurement the qubit is in a state with coherence $\mathcal C=2|\alpha\beta^*|$. However, in realistic situations such flexibility may not be possible, i.e. we may not be able to access all the necessary degrees of freedom of $S$ and we will have to split it into an accessible part and an inaccessible one. How much coherence can we expect to recover in that case?

Motivated by the importance of the question, we now compute how much coherence is recoverable on average when $\Q$ is immersed in an environment of which we can only access a subspace. The main difference with the example given above, is that $\mathcal Q$ and the environment probed by our measurement are generally not in a pure state such as the one in Eq.~\eqref{example}.
To do this, we consider a random pure state of a qubit in an $AK$-dimensional environment with Hilbert space $\mathcal A\otimes\mathcal K$ which is split into an accessible subspace of dimension $A$ and an inaccessible one of dimension $K$. Such a pure state is therefore sampled uniformly in a Hilbert space $\mathcal Q\otimes\mathcal A\otimes\mathcal K$ of dimension $2AK$. After tracing away the inaccessible environment $\mathcal K$, we are left with a $2A$-dimensional state in $\mathcal Q\otimes \mathcal A$ which can always be written as a $2A\times2A$ density matrix:
\begin{align}
\label{randomMatrix}
\hat \rho=
\begin{pmatrix}
\hat R_0&\hat X\\
\hat X^\dagger&\hat R_1
\end{pmatrix}
\end{align} 
where $\mathrm{Tr}(\hat R_0)$ and $\mathrm{Tr}(\hat R_1)$ are the probabilities of measuring the qubit in the alternatives $|0\rangle$ and $|1\rangle$ and $\hat X$ is the cross-term. The largest coherence of the qubit that we can obtain by optimally measuring the accessible space $\mathcal A$ is given by twice the trace norm of the cross-term: $\mathcal C=2\mathrm{Tr}|\hat X|$, see Eq.~\eqref{tracenorm} \cite{englert2000quantitative}. We recall that the trace norm of $\hat X$ can be computed as the sum of the square root of the $A$ eigenvalues of the matrix $\hat X^\dagger \hat X$, i.e.
\begin{align}
\mathrm{Tr}|\hat X|=\sum_{i=1}^{A}\sqrt{\lambda_i(\hat X^\dagger \hat X)}.
\end{align}

\begin{figure}[h!]
\begin{center}
\includegraphics[width=0.8\columnwidth]{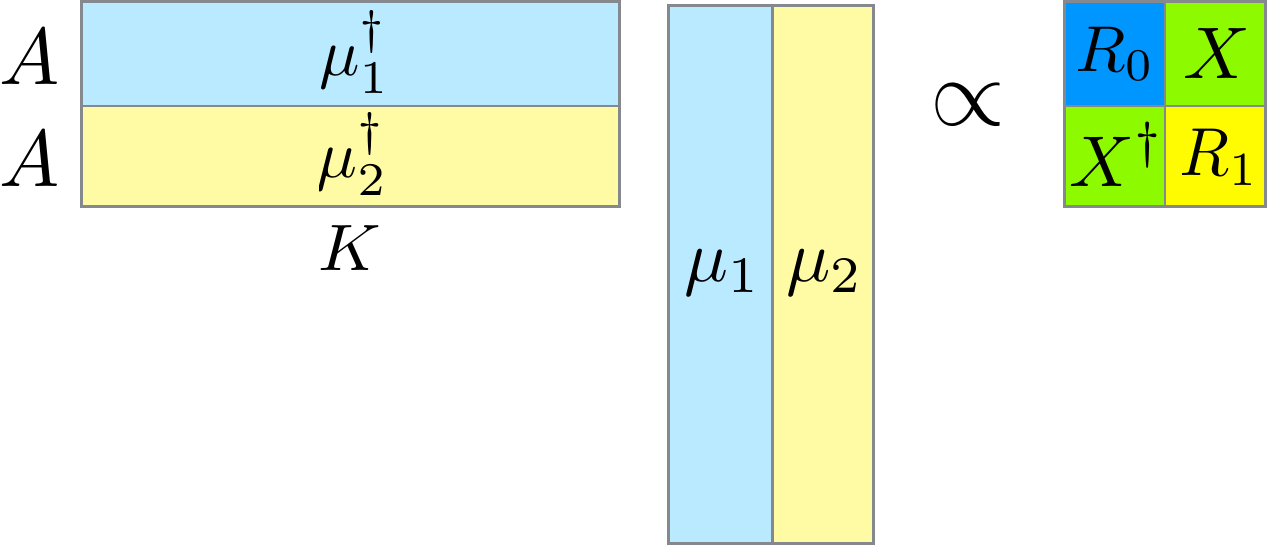}
\caption{\label{matrices}The $A\times A$ cross-term $X$ in the random matrix $\rho$ of Eq.~\eqref{randomMatrix} and \eqref{ginibre} is proportional to the product between two independent random matrices $\mu_1$ and $\mu_2$ that make up $\mu$.}
\end{center}
\end{figure}

The random $2A$-dimensional states $\hat\rho$ are statistically distributed according to the induced trace measure $P_{2A,K}(\hat \rho)$ and constitute a Ginibre ensemble \cite{KarolBook}. A way of sampling uniformly from such ensemble is to generate a $2A\times K$ complex gaussian random matrix $\mu$ (with entries sampled from the complex normal distribution centred on the origin and with unit variance) and then build the $2A\times2A$ density matrix 
\begin{align}
\label{ginibre}
\hat\rho=\frac{\mu^\dagger\mu}{\mathrm{Tr}(\mu^\dagger\mu)}.
\end{align}
However, when we calculate $\mathcal C$ we don't need the whole matrix $\hat\rho$, but only the $A\times A$ off-diagonal block $\hat X$ which is proportional to the product $M=\mu_1^\dagger\mu_2$ of two independent $A\times K$ complex gaussian random matrices $\mu_1$ and $\mu_2$ (see Fig.~\ref{matrices}). We find the proportionality factor by recalling that we are averaging over the whole ensemble and that the mean is linear, so we can take the average value of the denominator in \eqref{ginibre}: $\langle\mathrm{Tr}(\mu^\dagger\mu)\rangle=4AK$.
To find the average of $\mathrm{Tr}|M|$, we use the moments $m_\ell$ of the marginal distribution of eigenvalues of $M$. In particular, the average square root of the eigenvalues of $M$ is the moment of order $\ell=1/2$, i.e. $\langle\mathrm{Tr}|M|\rangle=Am_{\frac{1}{2}}$. We can compute such a moment by applying Eq.~(57) of Ref.~\cite{Akemann} to our matrices, and we find:
\begin{align}
\label{moment}
m_{\frac{1}{2}}=\frac{ 4\pi ^{5/2} (-1)^K \,
   _4\tilde{F}_3\left({\frac{1}{2},1-A,1-A,1-K\atop \frac{1}{2}-A,\frac{1}{2}-A,\frac{1}{2}-K}\bigg|
   1\right)}{A!\,\Gamma (A) \Gamma (K)}
\end{align}
where the function $_4\tilde{F}_3$ is a regularized Hypergeometric function. It is now straightforward to obtain the final result:
\begin{align}
\langle\mathcal C\rangle &= 2\langle\mathrm{Tr}|\hat X|\rangle=2\frac{\langle\mathrm{Tr}|\hat M|\rangle}{4AK}=\frac{m_\frac{1}{2}}{2K}.
\label{solution}
\end{align}

\section{examples}
\begin{figure}[h!]
\begin{center}
\includegraphics[width=0.9\columnwidth]{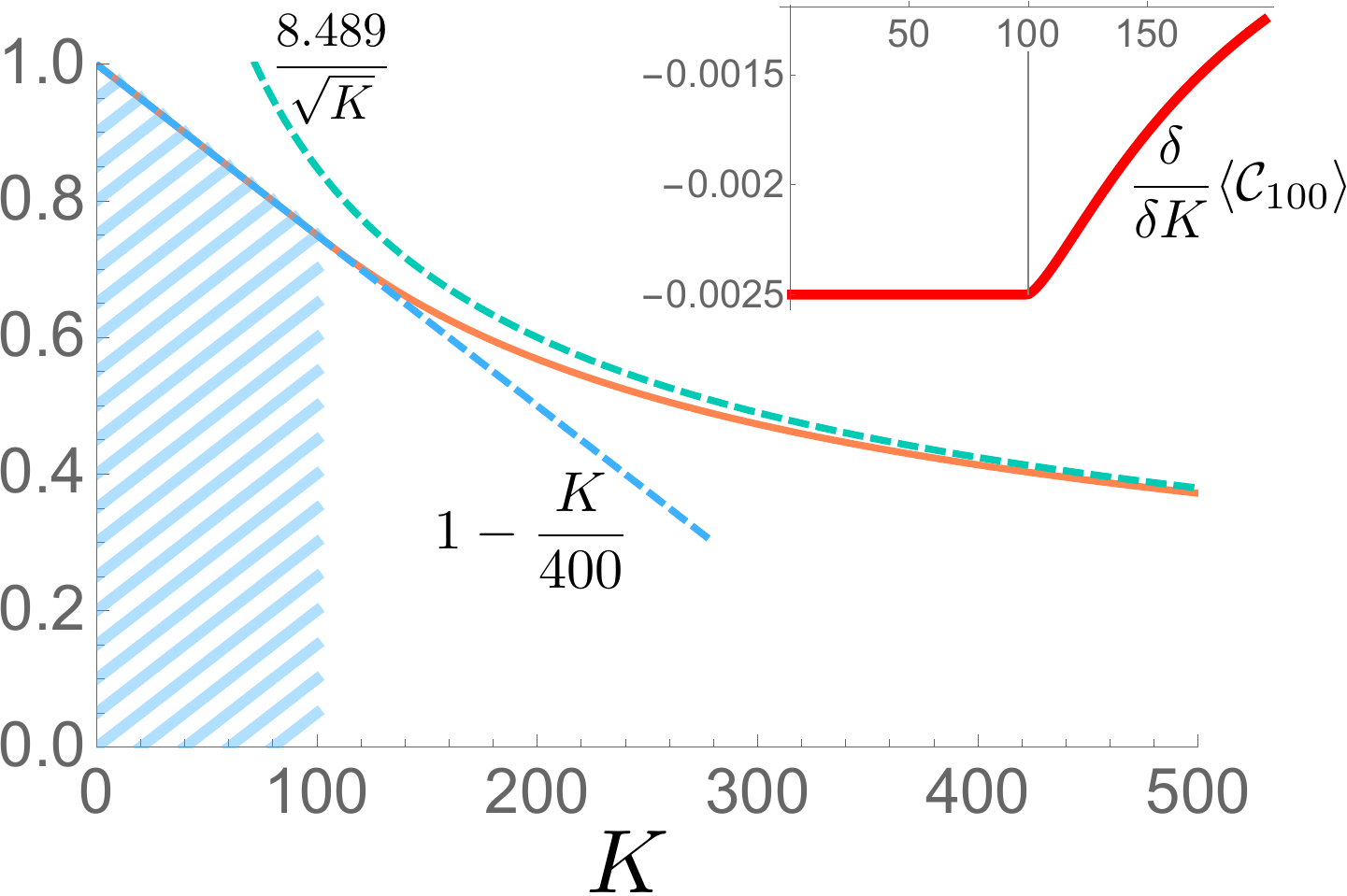}
\caption{\label{linearization}The average coherence $\langle\mathcal C\rangle$ for $A=100$ (solid orange), together with the high-$K$ and low-$K$ approximations (dashed). Up to $K=A$, the behaviour of $\langle\mathcal C\rangle$ is purely linear ($1-\frac{K}{4A}$, see inset). For $K>A$ the behaviour changes dramatically, and is asymptotic to $O(1/\sqrt{K})$. The shading indicates the linear region.}
\end{center}
\end{figure}

An expression with regularized Hypergeometric functions like Eq.~\eqref{moment} can be rather obscure. For this reason, we evaluate it explicitly for some values of $A$ and we show the two distinct behaviours that it exhibits, for $K\rightarrow\infty$ and for $1\leq K \leq A$.
Let's pick a uniformly random state of a qubit immersed in an entirely inaccessible $K$-dimensional environment. What is the average coherence of the qubit? As we are not performing any operation on the environment, this is equivalent to setting $A=1$ in Eq.~\eqref{moment} and \eqref{solution} and the answer is
\begin{align}
\langle\mathcal C_1\rangle=\frac{\pi ^{3/2} (-1)^K}{2 K! \Gamma \left(\frac{1}{2}-K\right)}\sim\frac{\sqrt{\pi }}{2 \sqrt{K}}\ \ (\mathrm{as\ }K\rightarrow\infty)
\end{align}
In other words, decoherence in absence of any intervention scales like $O(1/\sqrt{K})$ at a rate of $\sqrt{\pi}/2$. 
By controlling a two-dimensional space (i.e. $A=2$) we obtain
\begin{align}
\langle\mathcal C_2\rangle=\frac{\pi ^{3/2} (-1)^K (13-22K)}{32 K! \Gamma \left(\frac{3}{2}-K\right)}\sim\frac{11 \sqrt{\pi }}{16 \sqrt{K}}\ (\mathrm{as\ }K\rightarrow\infty)
\end{align}
By controlling a three-dimensional space we obtain
\begin{align}
\langle\mathcal C_3\rangle
&\sim\frac{107 \sqrt{\pi }}{128 \sqrt{K}}\quad (\mathrm{as\ }K\rightarrow\infty),
\end{align}
and so on. It turns out that the high-$K$ scaling is always $O(1/\sqrt{K})$. If instead we look at the scaling for $K\rightarrow0$ we find a linear behaviour: $\langle\mathcal C\rangle\sim1-\frac{K}{4A}$, the transition happening rather sharply at $A=K$ (see inset in Fig.~\ref{linearization}). We give an explicit example for $A=100$ in Fig.~\ref{linearization}.

\begin{figure}[h!]
\begin{center}
\includegraphics[width=\columnwidth]{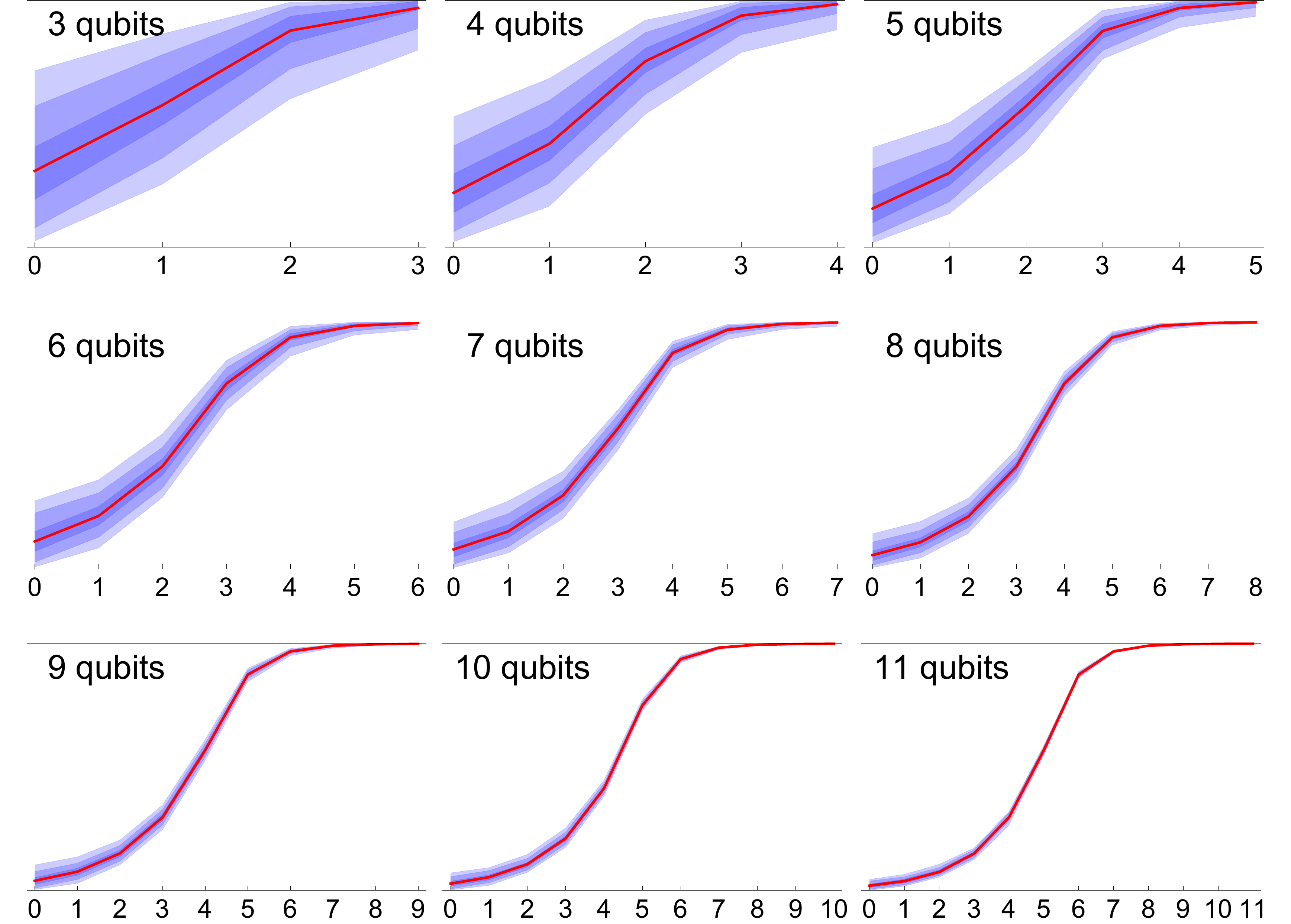}
\caption{\label{quantiles2}Graphs of the recoverable coherence $\langle \mathcal C\rangle$ of a qubit $\mathcal Q$ as a function of the number $a$ of qubits of the environment that we can control, where the total, $n$, is displayed on top. In blue we show the 50, 90 and 99 percentiles around the mean (red line) and all plots are from 0 to 1. We can see that $\langle \mathcal C\rangle$ transitions from a value close to 0 to a value close to 1 as we gain access to more than half of the environment qubits. Notice that as the total number of environment qubits grows, the mean becomes a better representative of the whole ensemble.}
\end{center}
\end{figure}

\begin{figure}[h!]
\begin{center}
\includegraphics[width=0.8\columnwidth]{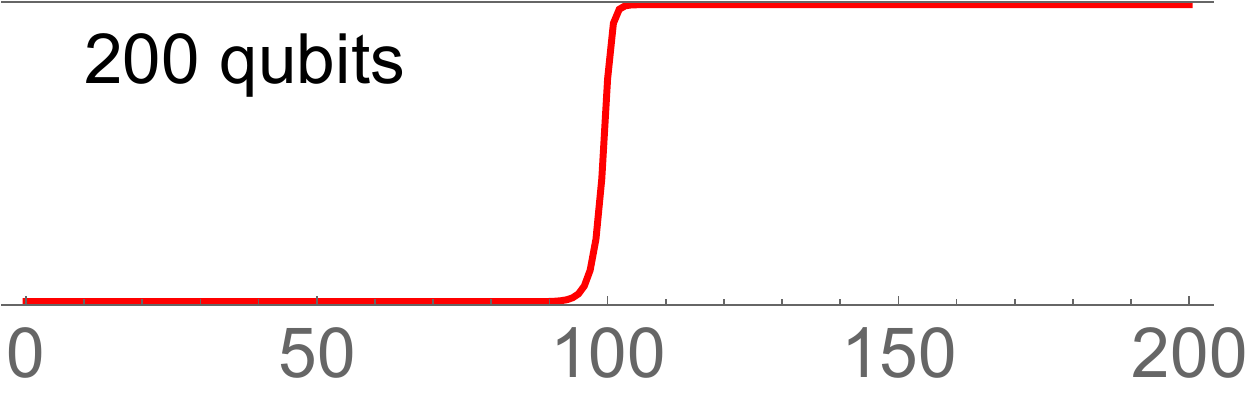}
\caption{\label{200qubits}The recoverable coherence $\langle\mathcal C\rangle$ of a qubit $\mathcal Q$ for an environment of $n=200$ qubits displays a very sharp increase from 0 to 1 as soon as one can control more than 100 of them, meaning that there exists an environment observable with elements $\{\hat\pi_j^\mathcal{A}\otimes\mathcal{\mathbb{\hat{1}}}^\mathcal{K}\}$ which project $\mathcal Q$ onto maximally coherent states.}
\end{center}
\end{figure}

We now give an example in terms of ensembles of qubits: let us consider our qubit $\mathcal Q$ immersed in an ensemble of $n$ other qubits, and let us pick a random state of all $n+1$ of them. We wish to compute how much coherence we can expect to recover on $\mathcal Q$ on average as we gain control of more and more qubits in the ensemble. In this case $A=2^a$ and $K=2^{n-a}=2^k$. We see that $\langle\mathcal C\rangle$ is close to zero for $a\lesssim k$ and it approaches 1 as $a \gtrsim k$ (see Fig.~\ref{quantiles2} and \ref{200qubits}). Note that such variation always happens across the same number of qubits. In fact, for $a\geq \ceil{n/2}$, the linear scaling makes it possible to compute the asymptotic behaviour $\langle\mathcal C\rangle\sim1-\frac{2^k}{2^{a+2}}$ (as $n\rightarrow\infty$). This law allows us to plot graphs for large numbers of qubits (see Fig.~\ref{200qubits}).

How typical is the value of $\langle\mathcal{C}\rangle$? To answer this question we produced thousands of random states from Ginibre ensembles of several qubits, for $n=3$ to $11$. The results are shown in Fig.~\ref{quantiles2}: already with an environment made of a handful of qubits, the average coherence is a typical value of the system, as all the states have a value of $\mathcal C$ that falls extremely close to $\langle\mathcal C\rangle$, the more so as $n$ grows.

\section{conclusions}
In this Letter we showed that quantum erasure can restore full coherence in a qubit even by addressing only about half of its environment. We also observed that the average recoverable coherence is a typical property of an ensemble of random states, i.e. $\mathcal C\sim \langle\mathcal C\rangle$ as $n\rightarrow\infty$. This means that the existence of such optimal measurement is almost always guaranteed. Our result is even more surprising if restated in terms of degrees of freedom, as the optimal measurement needs only address about $\sqrt{D}$ of the total number $D$ of degrees of freedom of the environment. Moreover, typicality assures us that this result holds for any such partition of the environment.

\section{Acknowledgements}
F.M.M. thanks Carlos Gonz\'alez Guill\'en for a fruitful consultation, Karol \.{Z}yczkowski for encouraging comments and Mike McDonell for illuminating discussions. This work was supported by the Canada Excellence Research Chairs (CERC) Program, the Natural Sciences and Engineering Research Council of Canada (NSERC) and the UK EPSRC.

\bibliography{bibliography}
\bibstyle{unsrt}

\end{document}